\documentclass[letterpaper,superscriptaddress,longbibliography,showpacs]{revtex4-1}
\usepackage[USenglish]{babel}
\usepackage{graphicx}
\usepackage{hyperref}

\usepackage{booktabs}
\usepackage{array}
\usepackage{longtable}
\usepackage{dcolumn}
\usepackage{float}
\usepackage{amsmath}
\usepackage{amssymb}
\usepackage{esint}

\usepackage{aas_macros}

\newcommand{\hunit}{\mathrm{km \ s^{-1} \ Mpc^{-1}}}
\newcommand{\der}{\mathrm{d}}
\newcommand{\lcdm}{\mathrm{\Lambda CDM}}
\newcommand{\densm}{\Omega_{\mathrm{m}}}
\newcommand{\densl}{\Omega_{\mathrm{\Lambda}}}
\newcommand{\densk}{\Omega_{\mathrm{k}}}
\DeclareMathOperator{\sinn}{sinn}

\begin{document}

\title{Constraints on the Dark Side of the Universe\\
and Observational Hubble Parameter Data}

\author{Tong-Jie Zhang}
\email{tjzhang@bnu.edu.cn}
\affiliation{Department of Astronomy, Beijing Normal University, Beijing 
100875, P.~R.~China}
\affiliation{Center for High Energy Physics, Peking University, Beijing 100871, 
P.~R.~China}
\author{Cong Ma}
\affiliation{Department of Astronomy, Beijing Normal University, Beijing 
100875, P.~R.~China}
\author{Tian Lan}
\affiliation{Department of Astronomy, Beijing Normal University, Beijing 
100875, P.~R.~China}

\begin{abstract}
This paper is a review on the observational Hubble parameter data that have 
gained increasing attention in recent years for their illuminating power on the 
dark side of the universe --- the dark matter, dark energy, and the dark age.  
Currently, there are two major methods of independent observational $H(z)$ 
measurement, which we summarize as the ``differential age method'' and the 
``radial BAO size method''.  Starting with fundamental cosmological notions 
such as the spacetime coordinates in an expanding universe, we present the 
basic principles behind the two methods.  We further review the two methods in 
greater detail, including the source of errors.  We show how the observational 
$H(z)$ data presents itself as a useful tool in the study of cosmological 
models and parameter constraint, and we also discuss several issues associated 
with their applications.  Finally, we point the reader to a future prospect of 
upcoming observation programs that will lead to some major improvements in the 
quality of observational $H(z)$ data.
\end{abstract}
\pacs{98.80.Es, 95.36.+x, 95.35.+d, 98.62.Ai, 98.62.Py, 98.65.Dx}

\maketitle

\section{Introduction}

The expansion of our universe has been one of the greatest attractions of 
scientific talents since the seminal work of Edwin Powell Hubble 
\citep{EPHubbleClassic} in 1929.  Hubble's compilation of observational 
distance-redshift (expressed in terms of radial velocity) data suggested a 
linear pattern of ``extra-Galactic nebulae'' (an archaic term for galaxies) 
receding from each other:
\begin{equation}
    \dot{\boldsymbol{x}} = H \boldsymbol{x},
\end{equation}
where $H$ is the proportional constant now bearing his name, and $x$ is the 
positional coordinates of a galaxy measured with our Galaxy as the origin.

The discovery of Hubble's Law marked the commencement of the era of 
quantitative cosmology in which theories of the universe can be subjected to 
observational test.  Since the days of Hubble, advances in technology have 
enabled astronomers to measure the light from increasingly deeper space and 
more ancient time, and our ideas of the entire history of the expanding 
universe have been gradually converging into a unified picture of Big 
Bang--Cold Dark Matter universe.  In this picture, the dominating form of 
energy density transited from radiation to dark matter, and relics of 
primordial perturbation were imprinted on today's observable CMB anisotropy and 
large-scale structures (LSS).  This picture is obtained from its two ends: the 
CMB last-scattering surface at $z \approx 1000$ and the LSS around us at $z 
\approx 0$.  The vast spacetime extent between both ends, in particular the era 
before reionization, remains mostly hidden from our view.  In addition, the 
past two decades' cosmological observations, especially those of type Ia 
supernovae (SNIa), indicated that the recent history of universal expansion is 
an acceleration, possibly driven by an unknown ``dark energy'' 
\citep{1998AJ....116.1009R,1999ApJ...517..565P} whose physical nature has not 
been identified.

Therefore it appears to us that our understanding of the universe is currently 
under the shade of three dark clouds --- the mysterious {\em dark energy} that 
drives late-time accelerated expansion, the nature of {\em dark matter} that is 
vital to the formation of structures, and the unfathomable {\em dark age} that 
has not yet revealed itself to observations.  This is the ``3-D universe'' in 
which possible answers to some of the most profound questions of physics are 
hidden.

In the face of these vast unknown sectors of the universe, any observational 
probe into its past history is invaluable.  Recently, the direct measurement of 
the expansion rate, expressed in terms of the Hubble parameter $H(z)$, is 
gaining increasing attention.  As a cosmological test, it can help with the 
determination of important parameters that affects the evolution of the 
universe, and reconstruct the history around key events such as the turning 
point from deceleration to acceleration.  As an observable, it manifests itself 
in various forms in different eras, especially in the baryon acoustic 
oscillation (BAO) features in the LSS that may be detectable in the dark age.  

This paper is a review on the current status of observational Hubble parameter 
data and its application in cosmology.  In Section \ref{sec:bg} we briefly 
review the cosmological background of an expanding universe.  In Section 
\ref{sec:ohd} we present two important observational methods of $H(z)$ 
observation, their principles and implementations.  Next, we review the 
important role of the observational $H(z)$ data in the study of cosmological 
models in Section \ref{sec:app}.  We will also discuss some issues associated 
with their application.  Finally, in Section \ref{sec:future}, we briefly 
discuss some ongoing efforts that promise possible improvements over the 
current status of $H(z)$ measurements.

\section{Background} \label{sec:bg}

In this section, we will review some basic ideas and definitions in cosmology 
that must be kept in mind in order to understand and interpret the 
observational $H(z)$ data and their implications.

\subsection{Spacetime, Metric, and Coordinates}\label{sec:coord}

The spacetime structure of the homogeneous, isotropic, and spatially flat 
universe is characterized by the Friedmann-Robertson-Walker (FRW) metric
\begin{eqnarray}
    \label{eq:frwmetric}
    g_{\mu\nu} =
    \left(
	\begin{array}{cccc}
	    -1 & & & \\
	      &  a^2(t) & & \\
	      & &  a^2(t) & \\
	      & & &  a^2(t)
	\end{array}
    \right).
\end{eqnarray}
The presence of the scale factor $a(t)$ means that the spacetime is not 
necessarily static.  In reality, we know that the universe is expanding, and 
$a(t)$ increases with time.

Using the metric (\ref{eq:frwmetric}), the infinitesimal spacetime interval 
scalar $\der s^2 = \der x_{\nu} \der x^{\nu} = g_{\mu\nu} \der x^{\mu} \der 
x^{\nu}$ is obviously
\begin{equation}
    \label{eq:interval_cart}
    \der s^2 = -c^2 \der t^2 + a^2(t) \der x^i \der x^i.
\end{equation}
Here we have used the four-coordinate vector $x^{\alpha} = (ct, 
\boldsymbol{x}^i)$ that has the dimension of length.

It is often useful to express the spatial components of the four-coordinate 
vector, i.e.~the ``comoving position'', in dimensionless spherical coordinates 
$\boldsymbol{x}^i = (r, \theta, \phi)$ in order to extend the metric to 
non-flat situations, and give the scale factor the dimension of length.  Under 
this convention, the spacetime interval (\ref{eq:interval_cart}) can be 
re-written as
\begin{equation}
    \label{eq:interval_sphe}
    \der s^2 = -c^2 \der t^2 + a^2(t) \left(\frac{\der r^2}{1 - k r^2}
    + r^2 \der \theta^2
    + r^2 \sin^2 \theta \der \phi^2\right)
\end{equation}
where $k$ is one of $\left\{-1, 0, 1\right\}$.  The parameter $k$ is the sign 
of the spatial curvature, and $k = 0$ if the universe is spatially flat.

We can further transform equation (\ref{eq:interval_sphe}) by introducing the 
coordinate
\begin{equation}
    \label{eq:defrcd}
    \chi = \int^r_0 \frac{\der r'}{\sqrt{1 - k r'^2}} = \sinn^{-1} r
\end{equation}
where the $\sinn$ function is a shorthand notation:
\begin{eqnarray}
    \sinn x =
	\begin{cases}
	    \sin  x & \text{for } k = 1, \\
		  x & \text{for } k = 0, \\
	    \sinh x & \text{for } k = -1.
	\end{cases}
\end{eqnarray}
Switching to the spatial coordinates $(\chi, \theta, \phi)$, the interval $\der 
s^2$ can be written as
\begin{equation}
    \label{eq:interval_chi}
    \der s^2 = -c^2\der t^2 + a^2(t) \left[\der \chi^2 + \sinn^2 \chi 
    \left(\der \theta^2 + \sin^2 \theta \der \phi^2\right)\right].
\end{equation}

The physical interpretation of $\chi$ can bee seen by placing ourselves at the 
origin $r = 0$ and consider a distant, comoving photon emitter in our 
line-of-sight direction with the coordinate $r = r_e$.  Rotate the coordinates 
so that the direction of the emitter has $\theta = 0, \phi = 0$, we find
\begin{equation}
    \der s^2 = -c^2 \der t^2 + a^2(t) \frac{\der r^2}{1 - k r^2}
\end{equation}
along the line-of-sight.  Let $t_e$ be the time of photon emission and $t_0$ 
that of its reception.  Since light-like worldlines have $\der s^2 = 0$, we 
find, for the photon:
\begin{equation}
    \label{eq:interpretchi}
    \int^{t_0}_{t_e} \frac{c \der t}{a(t)} = \int^{r_e}_{0} \frac{\der 
    r}{\sqrt{1 - k r^2}} = \chi(r_e).
\end{equation}
Consider the integrand in the left-hand side of equation 
(\ref{eq:interpretchi}).  The line element $\der x = c\der t$ is the physical 
distance the photon has traveled during the time interval $\der t$.  But by 
dividing the physical distance by $a(t)$ we get the comoving distance, 
therefore $\chi$ can be interpreted as the total, integrated comoving distance 
between the emitter and us.  If the space is flat, this comoving distance is 
just the difference in the radial coordinate $\Delta r = r_e - 0 = r_e$.

Sometimes it is convenient to introduce the {\em conformal time}, or the {\em 
comoving horizon} $\eta$ as the time component of the four-coordinate.  The 
conformal time is defined as
\begin{equation}
    \eta(t) = \int^{t}_{0} \frac{\der t'}{a(t')}
\end{equation}
where we integrate from the ``beginning of time''.  Using $c\eta$ as the time 
component, the comoving four-coordinate can be written as a dimensionless 
vector $x^{\alpha} = (c\eta, \chi, \theta, \phi)$ and the FRW metric takes the 
form
\begin{eqnarray}
    \label{eq:frwmetricinconf}
    g_{\mu\nu} =
    a^2(\eta) \left(
	\begin{array}{cccc}
	    -1 & & & \\
	      & 1 & & \\
	      & & \sinn^2\chi & \\
	      & & & \sinn^2\chi\sin^2\theta
	\end{array}
    \right).
\end{eqnarray}

\subsection{Expansion, Redshift, and the Hubble parameter}\label{sec:hparm}

In the introduction we mentioned Hubble's Law discovered in 1929.  Hubble's 
original paper had profound impact upon the history of astrophysics and, to a 
greater extent, mankind's perception of the universe, but here we only take 
some time to appreciate two of his timeless insights.

At the end of his paper Hubble briefly discussed the possible mechanisms for 
``displacements of the spectra'' (i.e.~redshift, in modern terms) in the de 
Sitter cosmology model in which the expansion of the universe is dominated by a 
vacuum energy.  He pointed out the two sources of the redshift: the first being 
``an apparent slowing down of atomic vibrations'' and the other attributed to 
``a general tendency of material particles to scatter''.  In today's words, the 
first is the special-relativistic effect of Doppler shift caused by the 
peculiar motion of galaxies, and the latter the general-relativistic, {\em 
cosmological redshift} which is linked to the expansion of the comoving grid 
itself.  In the rest of this article we will see how these two effects arise in 
modern cosmology and end up in our observational figures.

Hubble also noted that his proportional law might be ``a first approximation 
representing a restricted range in distance'', therefore deviating from the 
pure de Sitter model in which the Hubble constant $H$ should indeed be constant 
everywhere and throughout the history.  This is exactly how we see it now.  In 
the contemporary context, we usually define the {\em Hubble parameter} $H$ to 
be the relative expansion rate of the universe:
\begin{equation}
    \label{eq:hparamdef}
    H = \frac{\dot{a}}{a},
\end{equation}
and its value is usually expressed in the unit of $\hunit$.  The Hubble {\em 
constant}, $H_0$, now officially refers to the current value of the Hubble 
parameter.

However, it is not apparent how this definition is related to observable 
quantities.  Therefore we have to relate equation (\ref{eq:hparamdef}) to 
physical observables such as the length, the time, and the redshift.

First, we note that the cosmological redshift $z$ at any time $t$ is related to 
the scale factor $a$.  Let $t_e$ be the time of a photon's emission by a 
distant source and $t_0$ the time of its reception by an observer ``here and 
now.''  The observed redshift $z$ of the source satisfies
\begin{equation}
    \label{eq:zanda}
    1 + z = \frac{a(t_0)}{a(t_e)}.
\end{equation}
Consider an observer who surveys various sources with different redshifts.  The 
ideal survey is assumed to complete instantly --- all the observations are done 
at exactly the same time instance $t_0$.  Of course this is not strictly true, 
but we do not expect the scale factor $a(t_0)$ to change ``too fast'', and we 
expect the redshift not to change too much during the temporal scale of our 
interest (i.e.~typical lifetime of humans or observation programs).  If we do 
allow $t_0$ to change however, we are led to the Sandage-Loeb test 
\citep{1962ApJ...136..319S,1998ApJ...499L.111L} that observes the drifting of 
redshift during a long period of time.  Recently, the variation in the apparent 
magnitude of stable sources over $t_0$ has also been proposed as a possible 
cosmological test \citep{2010arXiv1001.3975Q}.  To our best knowledge, no data 
have been produced using these methods by now, and the proposed observation 
plans usually require $\sim$10 years to yield meaningful results 
\citep{2007PhRvD..75f2001C,2010PhLB..691...11Z} (however, we note that the idea 
of ``real-time cosmology'' is gaining interest recently, as reviewed by 
\citet{2010arXiv1011.2646Q}).  In this paper we will not focus on these 
methods, and we therefore neglect the passing of $t_0$.

We therefore differentiate equation (\ref{eq:zanda}) with respect to $t_e$, 
setting $t_0$ as a constant:
\begin{equation}
    \label{eq:scalefderiv}
    \frac{\der a(t_e)}{\der t_e} = - \frac{a(t_0)}{(1+z)^2} \frac{\der z}{\der 
    t_e} = - \frac{a(t_e)}{1+z} \frac{\der z}{\der t_e}.
\end{equation}
Dividing both sides by $a(t_e)$ we immediately find
\begin{equation}
    \label{eq:hdiffz}
    H(z) = -\frac{1}{1 + z} \frac{\der z}{\der t_e}.
\end{equation}
In Section \ref{sec:agemethod}, we will see how equation (\ref{eq:hdiffz}) is 
useful in measuring $H(z)$ by observing passively evolving galaxies.

Another way to relate $H(z)$ to observable quantities is to use the notion of 
the comoving distance $\chi$ introduced in equation (\ref{eq:defrcd}).  Take 
the time derivative of equation (\ref{eq:interpretchi}), we find
\begin{equation}
    \label{eq:comdistderiv}
    \frac{\der \chi}{\der t_e} = - \frac{c}{a(t_e)}.
\end{equation}
On the other hand, equation (\ref{eq:scalefderiv}) tells us about another 
derivative $\der t_e / \der z$.  Therefore we can find the derivative of $\chi$ 
with respect to the redshift:
\begin{eqnarray}
    \frac{\der \chi}{\der z} = \frac{\der \chi}{\der t_e} \frac{\der t_e}{\der 
    z} &=& \frac{c}{a(t_e)} \frac{a(t_e)}{(1+z)} \frac{\der t_e}{\der a(t_e)} 
    \nonumber \\
    &=& \frac{ca(t_e)}{a(t_0)} \frac{\der t_e}{\der a(t_e)} \nonumber \\
    &=& \frac{c}{a(t_0)H},
\end{eqnarray}
that is,
\begin{equation}
    \label{eq:comdistredshift}
    \frac{\der\left[a(t_0)\chi\right]}{\der z} = \frac{c}{H(z)}
\end{equation}
(also see, for example \citep{2003ApJ...598..720S,2006ApJ...637..598B}, but 
beware of different notation conventions).  If an observable object spans the 
length $a(t_0)\Delta\chi$ along the line-of-sight in some redshift slice 
$\Delta z$, we can estimate $H(z)$.  But how do we find such objects, 
i.e.~``standard rods''?  The idea is not to use the length of a concrete 
object.  Instead, we explore the spatial distribution of matter in the universe 
and focus on its {\em statistical} features, such as the BAO peaks in the 
two-point correlation function of the density field.  This is another method 
for extracting $H(z)$ data from observations.  (The quantity $a(t_0)\chi$ can 
be seen as a distance measure.  It is closely related to the ``structure 
distance'' $d_S = a(t_0)r$ defined by \citet[][Chapter  8]{2008cosm.book.....W} 
that naturally arises in calculating the power spectrum of LSS.  From equation 
(\ref{eq:defrcd}) we can see that the structure distance is equivalent to 
$a(t_0)\chi$ if the space is flat, or if the object is not too far away.)

We remark that the derivation of $H(z)$ expressed in terms of the standard rod, 
equation (\ref{eq:comdistredshift}), is only part of the story, for we have 
only considered a standard rod placed in the line-of-sight direction.  The 
transversely aligned test body is related to another important cosmological 
measure, namely the angular diameter distance $D_A(z) = a(z) r(z)$.  In an 
expanding universe, the angle $\Delta \theta$ subtended by a distant source is
\begin{equation}
    \label{eq:angularsize}
    \Delta \theta = \frac{a(z)}{D_A(z)} \Delta r_{\bot} = \frac{a(t_0)}{(1 + z) 
    D_A(z)} \Delta r_{\bot},
\end{equation}
where $\Delta r_{\bot}$ is the transverse spatial span of the source measured 
in the difference of comoving coordinate $r$ 
\citep{1972gcpa.book.....W,1999astro.ph..5116H}.  Naturally, once the physical 
scale of BAO is known and the BAO signal measured, the corresponding angular 
diameter distance can also be used as a cosmological test.

A classical cosmological test is the Alcock-Paczy{\'n}ski (AP) test 
\citep{1979Natur.281..358A} that can be expressed as another combination of 
$H(z)$ and $D_A(z)$.  The observable of the AP test is the quantity $A(z) = 
\Delta z/(z \Delta \theta)$ of some extended, spherically symmetric sources, 
where $\Delta z$ is the difference in redshft between the near and far ends of 
the object, and $\Delta \theta$ the angular diameter.  By our equations 
(\ref{eq:comdistredshift}) and (\ref{eq:angularsize}) it can be expressed as
\begin{equation}
    A(z) = \frac{\Delta z}{z \Delta \theta} = \frac{1 + z}{z} D_A(z) H(z) 
    \frac{\Delta \chi}{\Delta r_{\bot}}.
\end{equation}
A well-localized object placed in a region not too far away from us (so the 
non-trivial spatial geometry can be neglected) will have $\Delta \chi \approx 
\Delta r_{\shortparallel}$, the difference in the comoving coordinate along the 
line-of-sight.  Furthermore, for a nearly spherical object the approximation 
$\Delta r_{\shortparallel} \approx \Delta r_{\bot}$ holds, and $A(z)$ is 
reduced to
\begin{equation}
    A(z) = \frac{1 + z}{z} D_A(z) H(z).
\end{equation}
Clearly it cannot constrain $H(z)$ or $D_A(z)$ separately, but a combination of 
both.  The AP test, in more modern context, is usually understood as a 
geometrical effect on the statistical distribution of objects instead of 
concrete celestial bodies (see 
\citep{1996ApJ...470L...1M,1996MNRAS.282..877B,2003PhRvL..90b1302M}, and also
\citep{2003ApJ...598..720S,2004ApJ...615..573M} where the BAO effects were 
explicitly treated in the analysis).

Another combination of $H(z)$ and $D_A(z)$ naturally arises in the application 
of BAO scales measured in the spherically averaged galaxy distribution, namely 
the distance measure $D_V$ \citep{2007MNRAS.381.1053P} defined by
\begin{equation*}
    D_V(z) = \left[\frac{c z (1 + z)^2 D_A^2(z)}{H(z)}\right]^{1/3}.
\end{equation*}
To break the degeneracy between $H(z)$ and $D_A(z)$ in $D_V(z)$, the full 
2-dimensional galaxy distribution must be used, with the correlation function 
conveniently decomposed into the line-of-sight and transverse components (see 
section \ref{sec:sizemethod}, but also see \citep{2008PhRvD..77l3540P} for 
another decomposition scheme).

\section{Hubble Parameter from Observations}\label{sec:ohd}

Equations (\ref{eq:hdiffz}) and (\ref{eq:comdistredshift}) are the bare-bone 
descriptions of two established methods for $H(z)$ determination: the 
differential age method and the radial BAO size method respectively.  Either 
has been made possibly only by virtue of state-of-the-art redshift surveys such 
as the Sloan Digital Sky Survey (SDSS) \footnote{\url{http://www.sdss.org/}}.  
In this section, we will review both methods and the data they produced.

\subsection{The Differential Age Method}\label{sec:agemethod}

As equation(\ref{eq:hdiffz}) suggests, to apply age-dating to the expansion 
history, we look for the variation of ages, $\Delta t$, in a redshift bin 
$\Delta z$ \citep{2002ApJ...573...37J}.  The aging of stars serves as an 
observable indicator of the aging of the universe, because the evolution of 
stars is a well-studied subject, and stars' spectra can be taken and analysed 
to reveal information about their ages.  However, at cosmological distance 
scales it is not practical to observe the stars one by one: we can only take 
the spectra of galaxies that are ensembles of stars, possibly of different 
populations.  Since different star populations are formed at drastically 
different epochs, it is important for us to identify galaxies that comprises 
relatively uniform star populations, and to look for more realistic models of 
star formation.

The identification of such ``clock'' galaxies and the observation of their 
spectra have been carried out for archival data \citep{2003ApJ...593..622J}, 
and surveys such as the Gemini Deep Deep Survey (GDDS) 
\citep{2004ApJ...614L...9M}, VIMOS-VLT Deep Survey (VVDS) and the SDSS 
\citep{2010JCAP...02..008S}.  In addition, high-quality spectroscopic data have 
been acquired from the Keck I telescope for red galaxies in galaxy clusters 
\citep{2010ApJS..188..280S}.  Among the galaxies being observed, special 
notices should be paid to the luminous red galaxies (LRGs).  LRGs are massive 
galaxies whose constituent star populations are fairly homogeneous.  They make 
up a fair proportion in the SDSS sample and, beyond serving as ``clocks'', also 
trace the underlying distribution of matter in the universe (albeit with bias).  
Therefore, they reveal BAO signature in the density autocorrelation function 
that is used as the ``standard rod'' in the size method.

The identification and spectroscopic observations of these galaxies have led to 
direct determinations of $H(z)$ in low and intermediate redshift ranges.  
\citet{2003ApJ...593..622J} first obtained a determination of $H(z) = 69 \pm 12 
\ \hunit $ at an effective redshift $z \approx 0.09$ by the differential age 
method.  The work was later expanded by \citet{2005PhRvD..71l3001S} who 
extended the determination of $H(z)$ to 8 more redshift bins up to $z \approx 
1.8$.  This dataset was brought up-to-date by \citet[Table 
2]{2010JCAP...02..008S}.  Recently, new age-redshift datasets for different 
galaxy velocity dispersion groups have been made available 
\citep{2010MNRAS.408..213C} from SDSS data release (DR) 7 LRG samples.  We will 
see how these data are used in the study of cosmology models in Section 
\ref{sec:app}.

One may wonder why we take the effort to calculate the age differences in 
redshift bins when the age (or lookback time) data themselves can also be used 
to test cosmological models.  Indeed, the absolute age has been very useful in 
the estimation of cosmological parameters 
\citep{1995ApJ...443L..69S,1996Natur.381..581D,1997ApJ...484..581S,2001ApJ...550L.133A}.
Nevertheless, precise age-dating with low systematic biases can be only carried 
out on a narrow selection of sources.  On the other hand, by taking the 
difference of the ages in narrow redshift bins, the systematic bias in the 
absolute ages can hopefully cancel each other \citep{2004MNRAS.349..240J}.  Of 
course, we are not gaining anything for nothing even if the systematics 
perfectly cancel, for the binning of data lowers the total amount of 
measurements we can have.

A further approximation is that the majority of stars in the galaxies are 
formed almost instantaneously, in a single 
``burst''\citep{1999MNRAS.305L..16J}, therefore the intrinsic spread of the 
measured age arising from a heterogeneous star formation history can be 
expected to be small when fitting the observed spectra to stellar population 
models (specifically the single-stellar population (SSP) model used in 
\citep{2005PhRvD..71l3001S} and \citep{2010JCAP...02..008S}).  However, recent 
developments in the study of the formation history of galaxies and their 
stellar populations have led us to re-consider the assumptions made in previous 
works.  For example, using galaxy samples selected from numerical simulations, 
\citet{2010MNRAS.406.2569C} have shown that the SSP assumption may contribute 
to the systematic bias that varies across redshift ranges (hence failing to 
cancel, and propagating into the differential ages), while models that take the 
extended star formation history into account can be used to reduce the errors 
on $H(z)$.

In addition to the complexities in the stellar populations in each galaxy, the 
heterogeneity of galaxies in the sample also contributes to the errors in 
$H(z)$ measurements.  In \citep{2010MNRAS.406.2569C}, new sample selection 
criteria have been proposed that could help with obtaining more homogeneous 
galaxy samples for future analyses.

\subsection{The Radial BAO Size Method}\label{sec:sizemethod}

In Section \ref{sec:hparm}, we mentioned that the ``standard rod'' we seek in 
the sky is not an actual object but a statistical feature.  Indeed, the 
physical sizes of distant celestial objects are usually poorly known.  Worse 
still, even the {\em apparent}, i.e.~angular, sizes of galaxies are ambiguous 
because galaxies do not show sharp edges, and they appear fuzzy in images.  It 
can be imagined that size measurements along the line-of-sight could only lead 
to more problems, because even the angular sizes cannot help us much in this 
case.  Therefore, identifying a statistical ``standard rod'' becomes a 
necessity.

In the study of LSS, correlation functions are a simple and convenient measure 
of the statistical features in the spatial distribution of matter in the 
universe.  (For an early yet important treatment of the topic in the context of 
galaxy surveys, see \citep{1973ApJ...185..413P}.  For an example of other 
statistics in the context of BAO, see \citep{2010ApJ...718.1224X}.)  The 
two-point autocorrelation (i.e.~the correlation of a density field with itself) 
function $\xi(\boldsymbol{r}_1, \boldsymbol{r}_2)$ is one of the most used 
member in the correlation function family.  It measures the relatedness of 
position pairs in the same density field:  the joint probability of finding two 
galaxies in volume elements $\der V_1$ and $\der V_2$ located in the 
neighborhood of spatial positions $\boldsymbol{r}_1$ and $\boldsymbol{r}_2$ 
respectively is
\begin{equation}
    \label{eq:autocorr}
    \der P_{12} = n^2 \left[1 + \xi(\boldsymbol{r}_1, \boldsymbol{r}_2)\right] 
    \der V_1 \der V_2
\end{equation}
where $n$ is the mean number density.  If we believe that our universe is 
homogeneous in a statistical sense (i.e.~that the probabilistic distribution, 
or ensemble, from which the densities anywhere in our particular instance of 
the universe is drawn, does not vary from one area of the universe to another), 
the autocorrelation function becomes a function of $\boldsymbol{r} = 
\boldsymbol{r}_1 - \boldsymbol{r}_2$ only.  If we further assumes the 
(statistical) isotropy of the universe, the direction of $\boldsymbol{r}$ 
becomes unimportant, and the autocorrelation is dependent on the magnitude of 
$\boldsymbol{r}$ only (that is, $\xi = \xi(r)$).  Actually, our assumption of 
homogeneity is unnecessarily strong if we only work with two-point statistics, 
and all we need is the homogeneity in the first two moments of the underlying 
ensemble.  Such an ensemble is known as a wide-sense stationary (WSS) one.

For a WSS ensemble, the famous Wiener-Khinchin theorem says that the 
autocorrelation and the power spectrum $P(\boldsymbol{k})$ form a Fourier 
transform pair:
\begin{eqnarray}
    \label{eq:autocorrandpspec}
    P(k) = P(\boldsymbol{k}) &=& \int \xi(\boldsymbol{r}) e^{i\boldsymbol{k} 
    \cdot \boldsymbol{r}} \der^3 r, \nonumber \\
    \xi(r) = \xi(\boldsymbol{r}) &=& \frac{1}{(2\pi)^3} \int P(\boldsymbol{k}) 
    e^{-i\boldsymbol{k} \cdot \boldsymbol{r}} \der^3 k.
\end{eqnarray}(Here we write the power spectrum as $P(k)$, independent of the 
direction of the wave vector $\boldsymbol{k}$, under the same assumption of 
statistical isotropy mentioned above, but see discussion about redshift 
distortion below.)  Therefore, either the power spectrum or the autocorrelation 
can serve as a statistical tool to reveal the information contained in the LSS.  
Methods of estimating $P(k)$ has been developed and the importance of the power 
spectrum emphasized \citep{1994ApJ...426...23F,2004MNRAS.347..645P}.  On the 
other hand, for BAO surveys the autocorrelation function is probably a more 
straightforward way of presenting the results and testing their significance, 
because the BAO scales manifest themselves as protruding features (``peaks'' or 
``bulges'') in $\xi(r)$.  Actually, an estimator to the autocorrelation, along 
with its variance, can also be conveniently constructed from survey data using 
pair counts between the survey and random fields \citep{1993ApJ...412...64L}.

Needless to say, the ``true'' autocorrelation of the ensemble can never be 
fully known, because we have only one realization of the random field which is 
{\em the} universe we live in.  However, {\em estimating} the autocorrelation 
still makes sense because for today's large and well-sampled surveys the 
assumption of ergodicity is valid, under which the statistics can be performed 
to infer knowledges about the underlying ensemble \citep[Chapter 8 and Appendix 
D]{2008cosm.book.....W}.

Thus, if a random process induces some features in the spatial distribution of 
matter, the autocorrelation can be numerically computed to reveal such features 
that are otherwise hidden in the seemingly stochastic distribution.  
Furthermore, if the mechanism and properties of this process is well understood 
and quantitatively modelled, parameter estimation using these features becomes 
a possibility.

One of such possibility is provided by the BAO signatures in the LSS.  The 
mechanism of BAO effects must be traced back to the early universe before 
recombination, when the Compton scattering rate was much higher than the cosmic 
expansion rate.  Under this extreme limit, the tightly coupled photons and 
baryons can be treated as a fluid in which the perturbations drive sound waves.
The BAO effect in the cosmic microwave background (CMB) radiation has been 
subjected to extensive theoretical studies (see the early work of 
\citet{1970ApJ...162..815P}, a powerful analytical treatment by 
\citet{1995ApJ...444..489H} in Fourier space, another by 
\citet{2002PhRvD..65l3008B} in position space, and a review by 
\citet{2002ARA&A..40..171H}).  It has been confirmed and measured by CMB 
observations such as the Wilkinson Microwave Anisotropy Probe (WMAP) 
\citep{2003ApJS..148..135H,2003ApJS..148..233P,2007ApJS..170..288H,2009ApJS..180..296N,2010arXiv1001.4635L}.  
We will not discuss CMB in detail, and mainly concern ourselves with the 
aftereffect of BAO, namely its imprints on the large-scale distribution of 
matter.

The imprints of BAO in the observable distribution of galaxies today was 
predicted in theory (see \citep{1998ApJ...495...29G,1999MNRAS.304..851M}, and 
note that these papers were mainly written in the language of $P(k)$ rather 
than $\xi(r)$).  They were first detected in SDSS data by 
\citet{2005ApJ...633..560E}.  In \citep{2007MNRAS.381.1053P}, BAO measurements 
were made for SDSS and 2dF survey data using the power spectrum, and the 
results were presented as a general test of cosmological models.  The usage of 
BAO signatures in the LSS as a probe of $H(z)$ was discussed in 
\citep{2003ApJ...598..720S} (see also 
\citep{2003ApJ...594..665B,2005ApJ...633..575S,2007ApJ...665...14S}).

The idea of using BAO scales may appear to be simple and straightforward by our 
description so far, but in reality the autocorrelation function is subjected to 
various distortion effects that must be accounted for.

First, galaxies are not comoving objects.  Their apparent redshifts are 
inevitably a combined effect of the cosmological redshift and peculiar 
velocities (which was once contemplated by E~.P.~Hubble, see Section 
\ref{sec:hparm}).  Peculiar motion distorts the apparent correlation pattern in 
the redshift space and makes it anisotropic (see 
\citep{1983ApJ...267..465D,1987MNRAS.227....1K}).  Therefore, the isotropic 
autocorrelation function $\xi(r)$ fails to be a good measure.  In the 
literature the autocorrelation is usually expressed as a function of scales in 
the radial (line-of-sight) direction $\pi$ and transverse direction $\sigma$: 
$\xi = \xi(\sigma, \pi)$ with $r = \sqrt{\sigma^2 + \pi^2}$.  The observed 
$\xi(\sigma, \pi)$ will be a convolution between $\xi(r)$ and the peculiar 
velocity field.

Second, geometry of the spacetime also distorts the correlation pattern as the 
observation goes into deeper distances, where the spacetime geometry becomes 
non-trivial \citep{2000ApJ...528...30M}.  This is not a major concern for the 
analyses we will review in the rest of this section, because the survey data 
were from our local section of the universe ($z \approx 0$), and for $H(z)$ 
measurements only some thin slices in the redshift space were used.  However, 
future work that deals with deep survey data must take the geometrical 
distortions into analysis.

There is also the more delicate issue of biasing, meaning that the correlation 
pattern of the observed ``indicators'' does not necessarily reflect that of the 
underlying matter distribution \citep{1984ApJ...284L...9K}.  Among the effects 
contributing to the bias, the magnification effect by weak lensing is worthy of 
notice for our discussion, because it has a large effect on the radial 
autocorrelation function \citep{2007PhRvD..76j3502H,2008PhRvD..77f3526H}.

Using SDSS LRG samples in the redshift range $0.16 \le z \le 0.47$, BAO 
signature was detected in $\xi(\sigma, \pi)$ by \citet{2008ApJ...676..889O}.  
In their work the magnification bias by weak lensing was neglected, but in the 
redshift range it contributes little to the spherically averaged 
autocorrelation $\xi_0$ \citep{2007PhRvD..76j3502H}, also known as the 
monopole:
\begin{equation}
    \label{eq:baomonopole}
    \xi_{0}(r) = \frac{1}{2} \int_{-1}^{1} \xi(\sigma, \pi) \der \mu,
\end{equation}
where $r = \sqrt{\sigma^2 + \pi^2}$, and $\mu = \pi / r$.  In 
\citep{2008ApJ...676..889O} the BAO peak was detected in the monopole 
significantly, while the ridge-like BAO feature was weak in the anisotropic 
$\xi(\sigma, \pi)$.

\begin{figure}
    \includegraphics[width=0.8\textwidth]{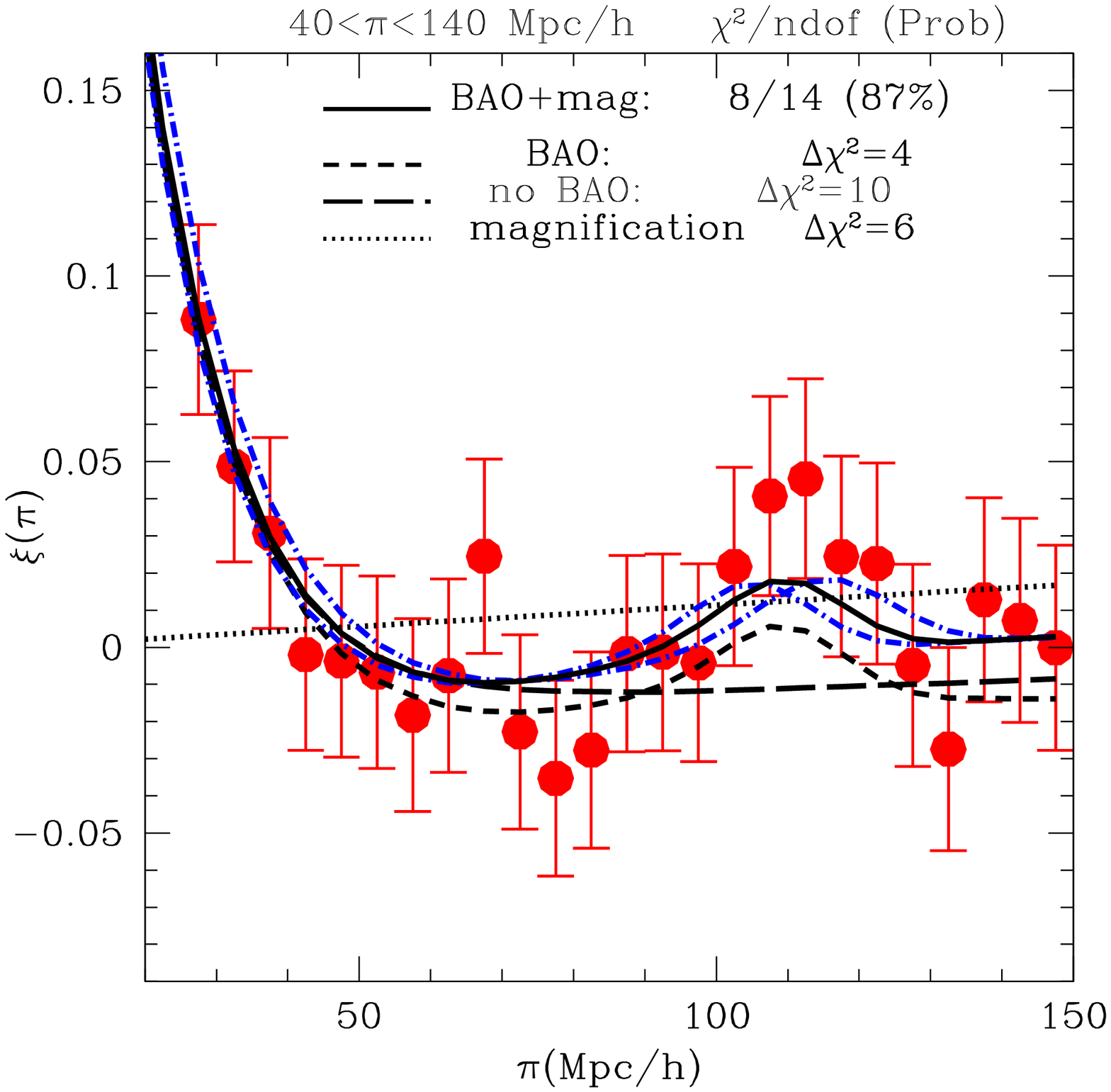}
    \caption{\label{fig:baodetection}
    Detection of radial ($\pi$-direction) BAO by \citet[Figure 
    13]{2009MNRAS.399.1663G} in the full LRG sample.  This is the correlation 
    pattern along the $\pi$-direction, and should not be confused with the 
    monopole pattern in Figure 3 of \citep{2008ApJ...676..889O}.  The effect of 
    weak lensing magnification can bee seen by comparing the solid and short 
    dashed curves, which shows that the magnification systematically moves the 
    peak location towards the higher scales.  The dash-dotted (blue) curve 
    shows the $1\sigma$ range by allowing the fiducial distance-redshift 
    relation used in the analysis to vary in a parameterized way, accounting 
    for the systematic error introduced by the mere using of a fiducial model.}
\end{figure}

Using improved LRG samples from SDSS DRs 6 and 7, and by modelling the weak 
lensing magnification bias, radial BAO detection and $H(z)$ measurements were 
made in redshift slices $z = 0.15 \sim 0.30$ and $z = 0.40 \sim 0.47$ by 
\citet{2009MNRAS.399.1663G} (see Figure \ref{fig:baodetection} for a 
presentation of the BAO detection).  Because these redshift slices were well 
separated, the two measurements were independent from each other.  (In previous 
works such as \citep{2007MNRAS.381.1053P} the samples overlapped and the 
results at different $z$'s were correlated.)

These $H(z)$ measurements were the first implementation of the radial BAO 
method.  Due to the distortion effects, confirming the significance of the 
baryon ridge detection becomes a demanding process, since each distortion 
effect has to be carefully modelled.  However, exact modelling of all the 
distortion effects on all scales is difficult, and when such modelling cannot 
be done exactly, these effects introduces systematic errors in the measurement 
of the BAO ridge's scale.

Despite these, the radial BAO size method still surpasses the age method in 
precision.  In fact, the combined statistical and systematic uncertainties 
presented an precision of $\sim$4$\%$ in $H(z)$ \citep[Table 
3]{2009MNRAS.399.1663G}.  This is intuitively perceptible.  As we have seen in 
Section \ref{sec:agemethod}, the age method is affected by the (possibly very 
large) systematic errors in age determination.  Since we can measure spatial 
quantities of galaxies, i.e.~the distribution of their positions, with much 
greater accuracy than we can do with temporal quantities related to some 
vaguely defined event (namely the time duration from star formation in the red 
galaxies to now), one may intuitively expect lower uncertainties from the 
radial size method than the differential age method.

A subtle issue of possible circular logic in the analysis also contributes to 
the systematic errors in this method.  In \citep{2009MNRAS.399.1663G}, a 
fiducial flat $\lcdm$ model and parameters were used to convert redshifts into 
distances, and to gauge the comoving BAO scales in the selected redshift slice, 
$r_{\mathrm{BAO}}$ to that of the CMB measured by 5-year WMAP, 
$r_{\mathrm{WMAP}} = 153.3 \pm 2.0 \mathrm{Mpc}$ (see 
\citep{2009ApJS..180..330K}) to yield the estimation $H_{\mathrm{BAO}}(z)$:
\begin{equation}
    \label{eq:estwithfidmodel}
    \frac{H_{\mathrm{BAO}}(z)}{r_{\mathrm{BAO}}} = 
    \frac{H_{\mathrm{fid}}(z)}{r_{\mathrm{WMAP}}},
\end{equation}
where
\begin{equation*}
    H_{\mathrm{fid}}(z) = H_0 \sqrt{\densm (1 + z)^3 + (1-\densm)}
\end{equation*}
and $\densm = 0.25$ \footnote{Another way to present the measurement results 
for use in cosmological parameter constraint $\Delta z_{\mathrm{BAO}} = 
r_{\mathrm{BAO}}H(z)/c$.  Schematically, this is done by approximating the 
derivative in equation (\ref{eq:comdistredshift}) with a ratio of differences, 
and identifying the interval $a(t_0)\Delta\chi$ with the measured comoving BAO 
scale.  In Section \ref{sec:app} we briefly discuss its usage.}.  The use of a 
fiducial model introduces bias in all measurements, which is hard to model 
exactly, but an analysis of this effect was performed using Monte Carlo 
simulations so that its contribution to the systematic uncertainties could be 
assessed.  The authors of \citep{2009MNRAS.399.1663G} hence argued that the 
measurement results are model-independent, therefore is useful as a general 
cosmological test.  The reader may also consult \citep{2007MNRAS.381.1053P} for 
a different approach to this issue, using cubit spline fit of the 
distance-redshift relation so that the result could be applied to a large class 
of models without having to re-analyze the power spectra for each model to be 
tested.

\paragraph*{A Word on the Dispute over the Radial BAO Detection.}  Currently 
there is some dispute over the claimed detection of radial BAO and measurement 
of $H(z)$ in \citep{2009MNRAS.399.1663G}.  \citet{2009arXiv0901.1219M} argued 
against the methods in \citep{2009MNRAS.399.1663G} and the statistical 
significance of the claimed BAO detection.  \citet{2010ApJ...719.1032K} 
analyzed the SDSS DR7 sample of LRGs and obtained similar results to 
\citep{2009MNRAS.399.1663G}, but offered another interpretation using the 
$\chi^2/(\text{degree of freedom})$ statistic and the Bayesian evidence 
\citep{2009ARNPS..59...95L} that disfavors a statistically significant 
detection.  On the other hand, the recent research of 
\citet{2010arXiv1011.2481T} claims that the radial BAO feature is not a fluke, 
albeit certain assumptions made this re-assessment somewhat optimistic.  The 
authors of \citep{2009MNRAS.399.1663G} also defended their work in 
\citep{2010arXiv1011.2729C}.  We refer to these variety of arguments and 
opinions to remind the reader of these ongoing investigations.  Nevertheless, 
we believe that the general method of measuring $H(z)$ using radial BAO is 
well-motivated and promising regardless of its current implementation, as it is 
expected to give more definitive results of radial BAO and $H(z)$ measurement 
with upcoming redshift survey projects \citep{2010ApJ...719.1032K}.

\section{Observational Hubble Parameter as a Cosmological Test}\label{sec:app}

The efforts in obtaining observational $H(z)$ data was certainly done with the 
goal of testing cosmological models in mind.  In \citep{2003ApJ...593..622J} 
the observation $H(z)$ at $z \approx 0.09$ was used to constrain the equation 
of state parameter of dark energy.  In \citep{2005PhRvD..71l3001S} the 
redshift-variability of a slow-roll scalar field dark energy potential was 
constrained by the differential age $H(z)$ data.  The same dataset was also 
utilized in the study of the $\lcdm$ universe, especially the summed neutrino 
masses $m_{\nu}$, the effective number of relativistic neutrino species 
$N_{\mathrm{rel}}$, the spatial curvature $\Omega_{\mathrm{k}}$, and the dark 
energy equation of state parameter $\omega$ \citep{2008JCAP...10..038F}.  The 
updated $H(z)$ data presented in \citep{2010JCAP...02..008S} was used by their 
authors to improve the results obtained in earlier papers.

\begin{figure}
    \includegraphics[width=0.85\textwidth]{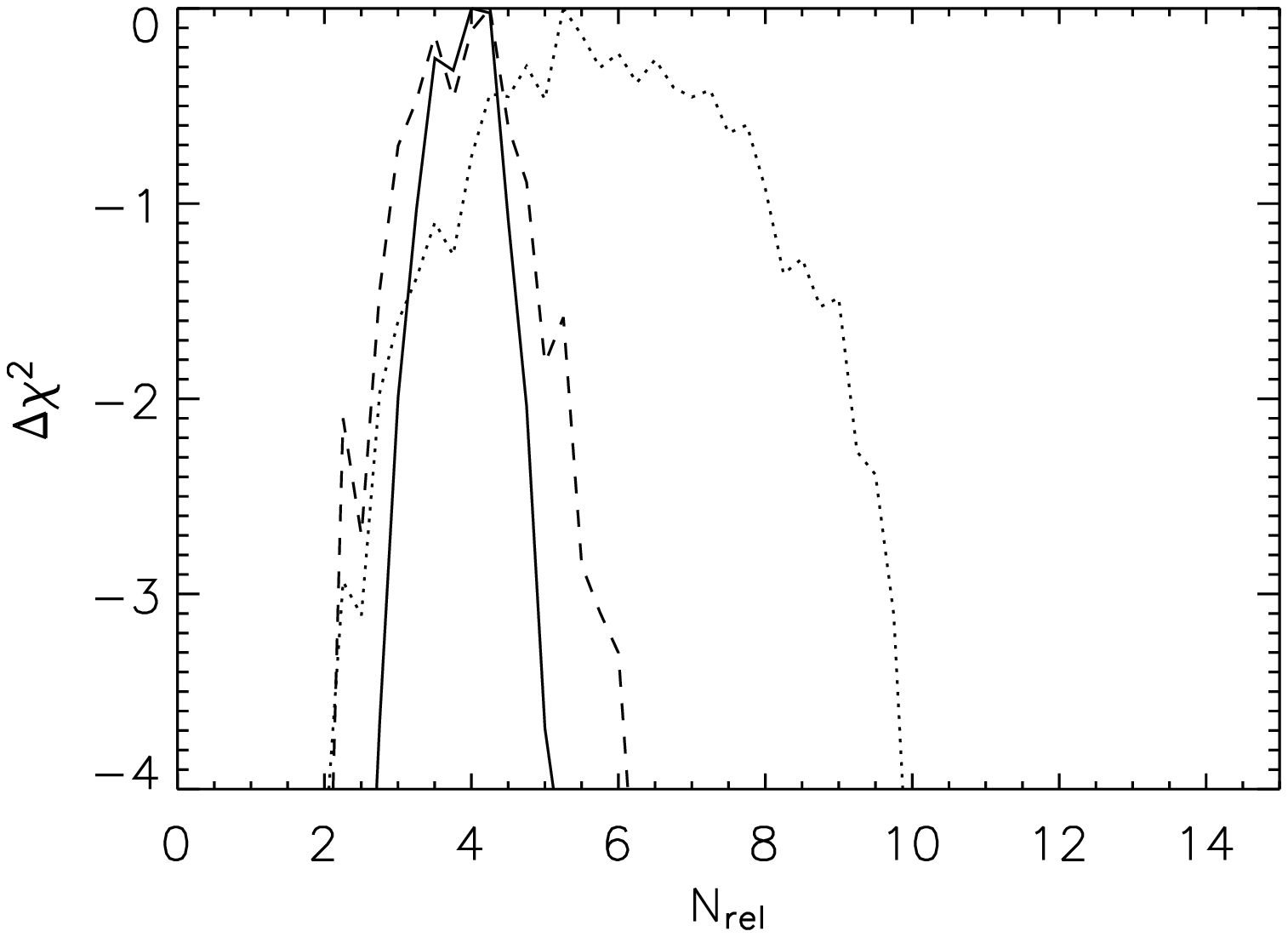}
    \caption{\label{fig:sternetalneff}Constraint on the effective number of 
    relativistic neutrino species, $N_{\mathrm{rel}}$, by 
    \citet{2010JCAP...02..008S} using their $H(z)$ measurements by the 
    differential age method.  Dotted line plots the 5-year WMAP 
    \citep{2009ApJS..180..306D} likelihood, dashed line plots the likelihood 
    with WMAP and $H_0$ determined by \citet{2009ApJ...699..539R}, and the 
    solid like the likelihood with WMAP, $H_0$ and $H(z)$ data.  Adding $H(z)$ 
    data helped refining the constraint to $N_{\mathrm{rel}} = 4\pm0.5$ at 
    1-$\sigma$ level.  The improvement in the constraint by adding $H(z)$ data 
    is evident.  Note that this figure displays the deviation of the $\chi^2$ 
    statistic from its minimum {\em inverted} ($\Delta \chi^2 = 
    \chi^2_{\mathrm{min}} - \chi^2$).  The intersections of the $\Delta \chi^2$ 
    plots with the constant $\Delta \chi^2 = 4$ line correspond to 2-$\sigma$ 
    constraints.}
\end{figure}

In particular, the combination of CMB and $H(z)$ observation is a very 
effective way to constrain $N_{\mathrm{rel}}$ \citep[see the reproduced Figure 
\ref{fig:sternetalneff} in this paper]{2010JCAP...01..003R}.  In this paper we 
will not go further into the topic of cosmic neutrinos, which is intrinsically 
related to fundamental physics.  However, we should point out a remarkable 
result, that the $H(z)$ data, when used jointly with CMB and other late-era 
cosmological tests, offer valuable insight into the neutrino properties related 
to the much earlier universe, independent of Big-Bang neucleosynthesis (BBN) 
\citep{2007ApJ...662...15I,2009PhR...472....1I} tests.  Moreover, the BBN 
constraints are obtained using Helium abundance measurements that are subjected 
to the systematic biasing effects arising from late-time neucleosysthesis.  
Therefore, $H(z)$ data is an important consistency check measure in the 
presence of this systematic uncertainty \citep{2010JCAP...01..003R}.

\begin{figure}
    \includegraphics[width=0.85\textwidth]{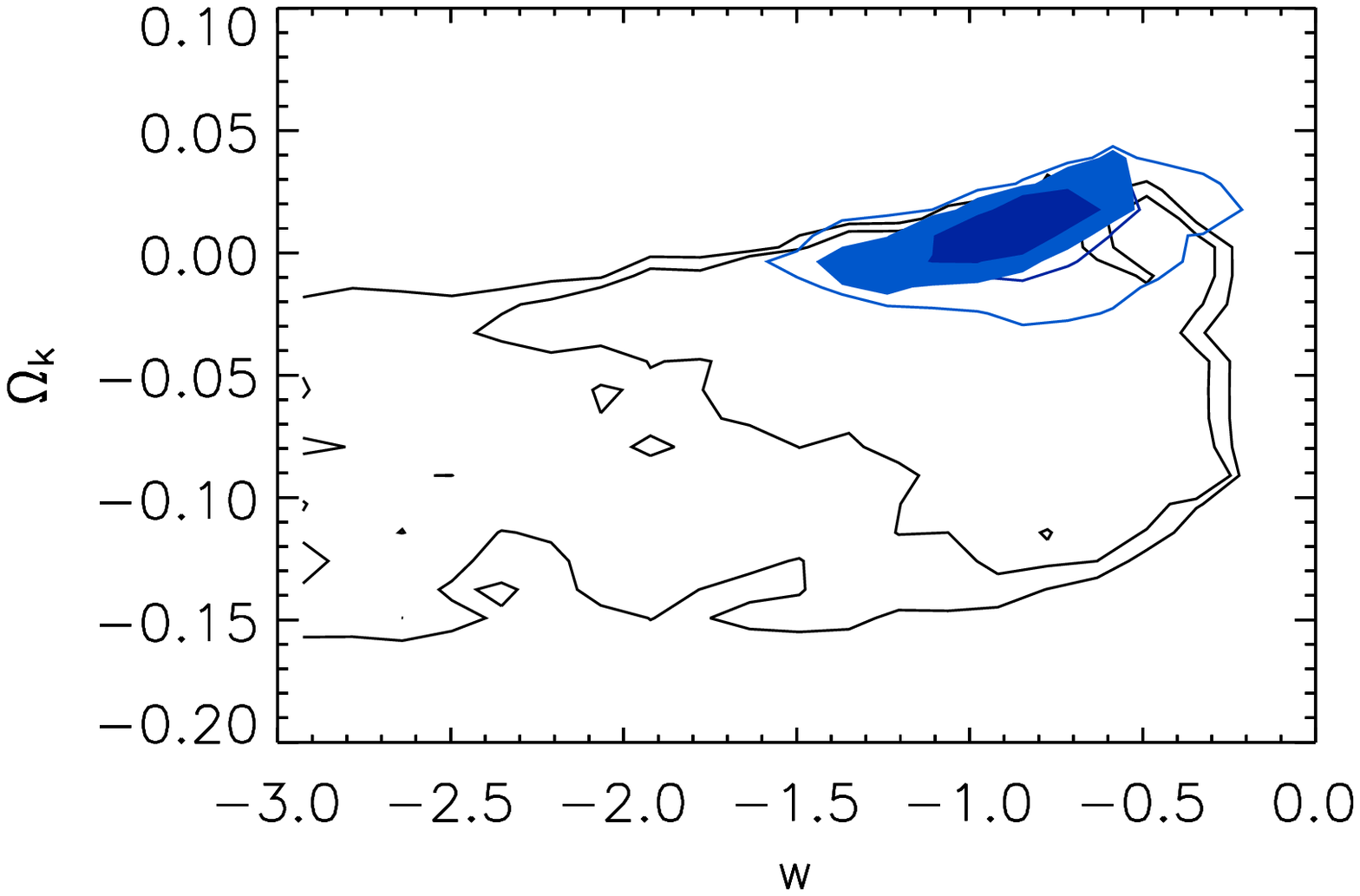}
    \caption{\label{fig:sternetalcurv}Joint constraint on the energy density 
    corresponding to the spatial curvature, $\densk$, and the dark energy 
    equation of state parameter, $w$, by \citet{2010JCAP...02..008S}.  The 
    large, irregular regions bounded by dark contours were from 5-year WMAP 
    alone.  The blue contours were obtained by adding $H_0$ constraints.  
    Filled regions were obtained by further adding $H(z)$ data.  The 
    application of $H(z)$ data helps with breaking the degeneracy between 
    $\densk$ and $w$.}
\end{figure}

Figure \ref{fig:sternetalcurv} shows that adding $H(z)$ data helps with 
breaking the degeneracy between spatial curvature and dark energy equation of 
state.  In the $\lcdm$ universe, both the dark energy and spatial curvature 
becomes dominant in recent epochs.  Therefore, separating their respective 
effects on the expansion of the universe becomes important, as well as 
challenging \citep{2007JCAP...08..011C,2009MNRAS.397..431V}.  While other tests 
using the combination of weak lensing and BAO are likely to measure the 
curvature distinctively in the future 
\citep{2006ApJ...637..598B,2009ApJ...690..923Z}, our current knowledge of 
$H(z)$ is still a valuable complement to other tests in the sense of 
DE-curvature degeneracy breaking \citep{2008JCAP...10..038F}.

The data produced by the BAO size method in \citep{2009MNRAS.399.1663G} is 
scarcer in quantity but of higher precision.  In \citep{2009MNRAS.399.1663G} 
they were extrapolated to $z = 0$ to offer an independent estimation of the 
Hubble constant $H_0$, and were used to test the accelerated expansion of the 
universe.  It has been demonstrated that the radial $\Delta z_{\mathrm{BAO}}$ 
measurements is able to put stringent constraints over the dark energy 
parameters \citep{2009PhRvL.103i1302G}.

In the papers cited above, the parameter constraints obtained from 
observational $H(z)$ data were shown to be consistent with other cosmological 
tests, such as the CMB anisotropy.  In this way, the observational $H(z)$ data 
presents themselves as a useful, {\em independent} cosmological test.  In 
particular, it serves as a powerful tool to break the degeneracy between the 
curvature and dark energy parameters.

\begin{figure}
    \includegraphics[width=0.8\textwidth]{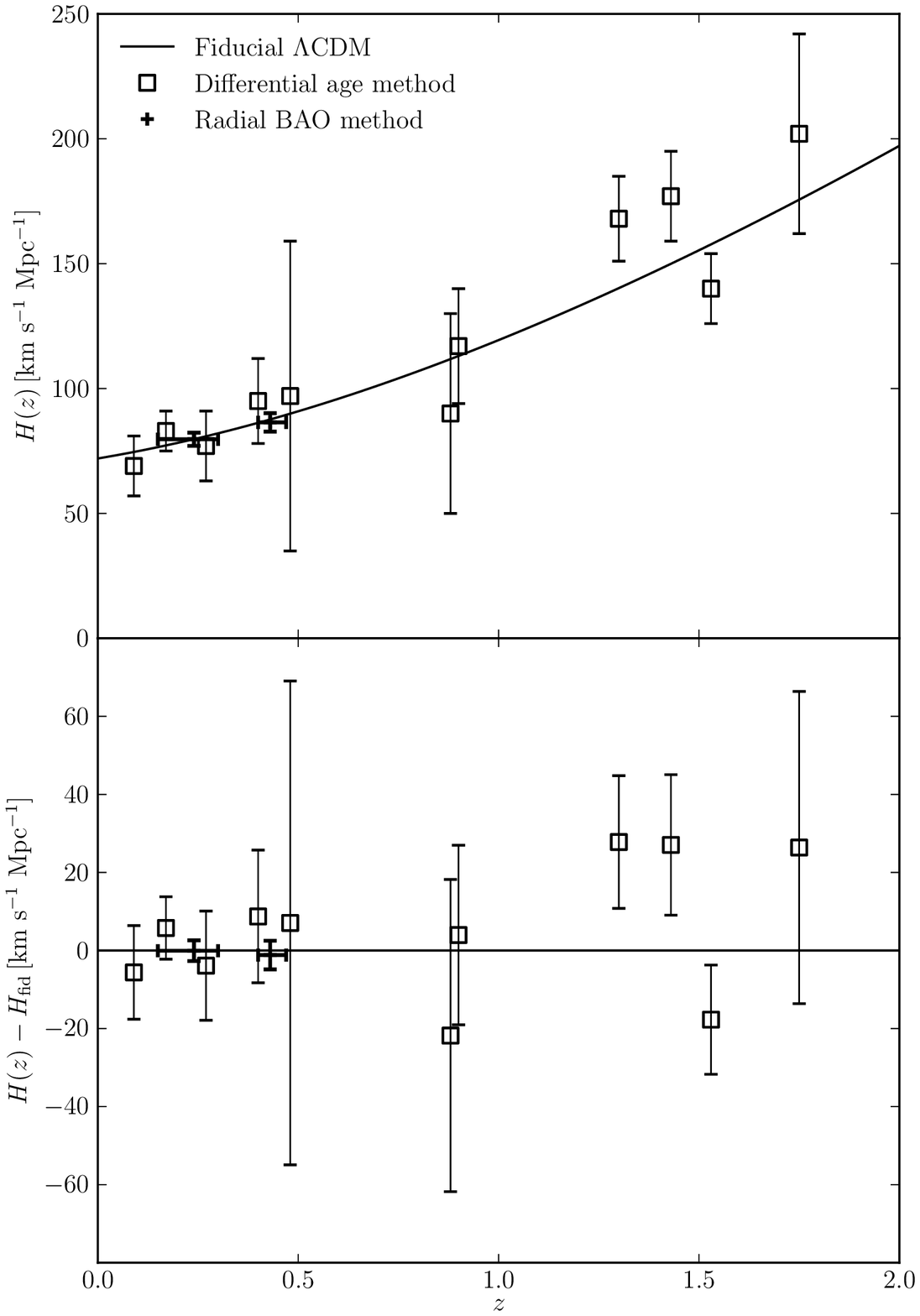}
    \caption{\label{fig:obsdata}Top panel --- the available $H(z)$ data from 
    both differential age method and radial BAO size method (see Table 
    \ref{tab:obsdata} and references therein).  The solid curve plots the 
    theoretical Hubble parameter $H_{\mathrm{fid}}$ as a function of $z$ from 
    the spatially flat $\lcdm$ model with $\densm = 0.25, \densl = 0.75, 
    \text{and } H_0 = 72\ \hunit$.  Bottom panel --- the same data, but the 
    residuals with respect to the theoretical model $H_{\mathrm{fid}}$ are 
    plotted.  In both panels, the $z$ error bars on the measurements from the 
    radial BAO method are used to mark the extents of the two independent 
    redshift slices in which the BAO peaks were measured.}
\end{figure}

These up-to-date data are summarized in Table \ref{tab:obsdata}.  In Figure 
\ref{fig:obsdata} we plot the $H(z)$ data versus the redshift.  To help 
visualizing the data, we also plot a spatially flat $\lcdm$ model with $\densm 
= 0.25, \densl = 0.75, \text{and } H_0 = 72\ \hunit$.

\begin{table}
    \caption{\label{tab:obsdata}The set of available observational $H(z)$ data}
    \begin{ruledtabular}
	\begin{tabular}{d D{,}{\,\pm\,}{-1} c c}
	    z & H(z),1\sigma\text{ error}\footnotemark[1]\footnotetext{$H(z)$ 
	    figures are in the unit of $\hunit$.} & References & Remarks  \\
	    0.09 & 69,12  & \citep{2003ApJ...593..622J,2010JCAP...02..008S} & 
	    \\
	    0.17 & 83,8   & \citep{2010JCAP...02..008S} & \\
	    0.24 & 79.69,2.65\footnotemark[2] & \citep{2009MNRAS.399.1663G} & 
	    In the redshift slice $0.15 \sim 0.30$    \\
	    0.27 & 77,14  & \citep{2010JCAP...02..008S} &   \\
	    0.4  & 95,17  & \citep{2010JCAP...02..008S} &   \\
	    0.43 & 86.45,3.68\footnotemark[2]\footnotetext{Including both 
	    statistical and systematic uncertainties: $\sigma =  
	    \sqrt{\sigma^2_{\mathrm{sta}} + \sigma^2_{\mathrm{sys}}}$.} & 
	    \citep{2009MNRAS.399.1663G} & In the redshift slice $0.40 \sim 
	    0.47$    \\
	    0.48 & 97,62  & \citep{2010JCAP...02..008S} &   \\
	    0.88 & 90,40  & \citep{2010JCAP...02..008S} &   \\
	    0.9  & 117,23 & \citep{2010JCAP...02..008S} &   \\
	    1.3  & 168,17 & \citep{2010JCAP...02..008S} &   \\
	    1.43 & 177,18 & \citep{2010JCAP...02..008S} &   \\
	    1.53 & 140,14 & \citep{2010JCAP...02..008S} &   \\
	    1.75 & 202,40 & \citep{2010JCAP...02..008S} &
	\end{tabular}
    \end{ruledtabular}
\end{table}

\begin{figure}
    \includegraphics{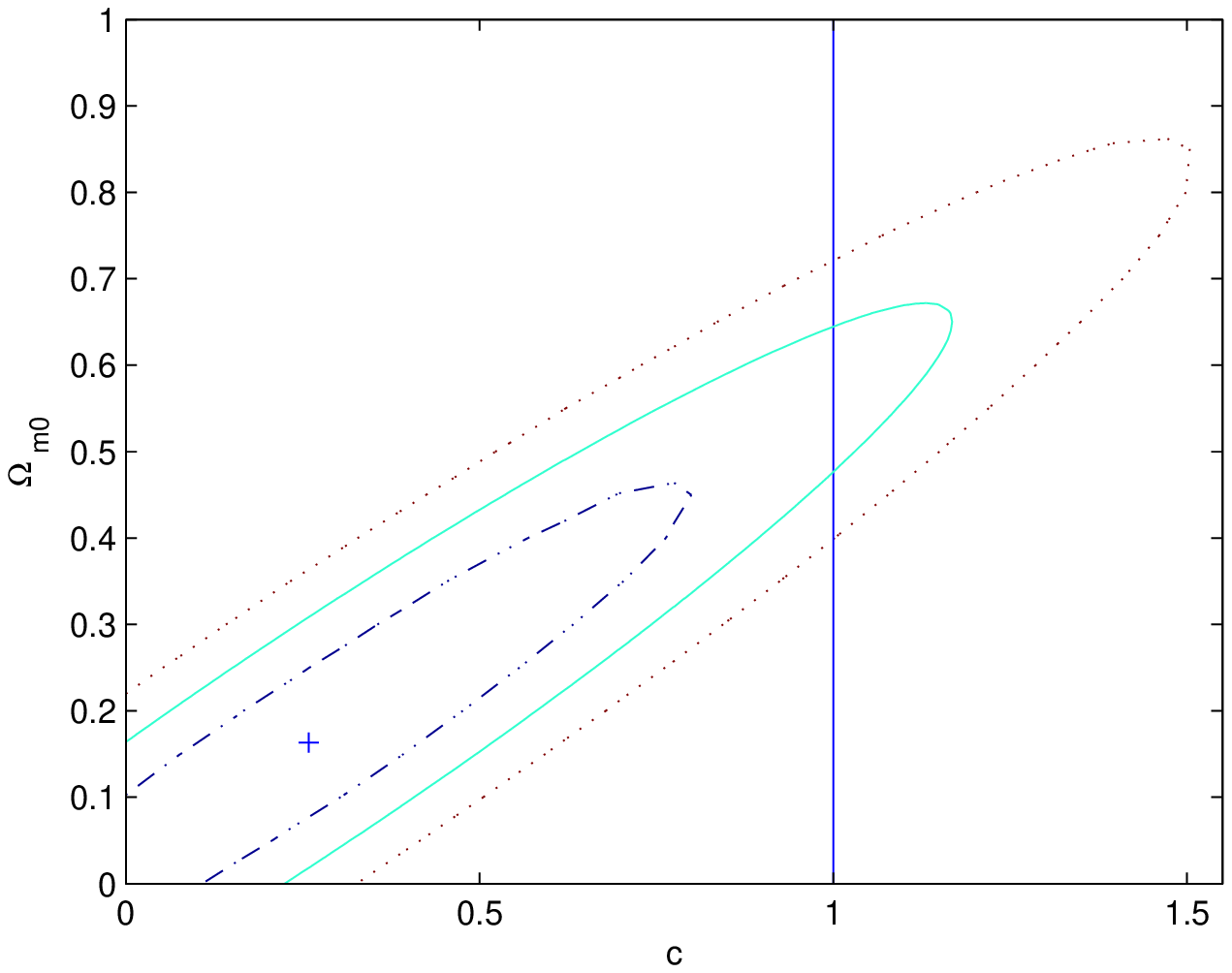}
    \caption{
    \label{fig:holo}
    Parameter constraints for the holographic dark energy model in the 
    $\densm$-$c$ plane, by \citet{2007MPLA...22...41Y}.  The constraints were 
    obtained using age-determined $H(z)$ data in \citep{2005PhRvD..71l3001S} 
    alone.  The cross in the lower-left marks the best-fit value.  The 
    dash-dotted, solid, and dotted contours marks the $68.3\%$, $95.4\%$, and 
    $99.7\%$ confidence regions respectively.  Although some degeneracy exists, 
    it is evident that the data favor models with $c < 1$.}
\end{figure}

In addition to the above authors, the observational $H(z)$ datasets have been 
widely used to put various cosmological models under test.  The first adopters 
included \citet{2007MPLA...22...41Y} and \citet{2006ApJ...650L...5S} who made 
use of the $H(z)$ results of \citep{2005PhRvD..71l3001S} in the study of dark 
energy.  In \citep{2007MPLA...22...41Y} the $H(z)$ data alone were used to 
constrain the parameters of the holographic dark energy model, especially th 
$c$ parameter that determines the dynamical history of the expanding universe 
(see Figure \ref{fig:holo}).  The same dataset has also been used to study 
modified gravity theory such as $f(R)$ gravity in the context of cosmology 
\citep{2008JCAP...09..008C}.  The updated data in \citep{2010JCAP...02..008S} 
and \citep{2009MNRAS.399.1663G} have been adopted to constrain the parameters 
in more exotic dark energy models, 
e.g.~\citep{2010JCAP...06..002X,2010JCAP...07..018D}.

Beyond parameter constraints, the observational $H(z)$ data are also applicable 
in non-parametric, model-independent cosmological tests.  For example, the $Om$ 
statistic by \citet{2008PhRvD..78j3502S}, defined by
\begin{equation}
    \label{eq:om}
    Om(z) = \frac{h^2(z) - 1}{(1+z)^3 - 1},
\end{equation}
where $h$ is the dimensionless Hubble parameter, $h = H(z)/H_0$.  This 
statistic is useful as a null test of dark energy being a cosmological
constant $\mathrm{\Lambda}$, and is more robust than parameterizations of the 
dark energy equation of state.  Another result for testing $\mathrm{\Lambda}$ 
that incorporates $H(z)$ data (the $\mathcal{L}_{\mathrm{gen}}$ test) is given 
by \citet{2008PhRvL.101r1301Z}, with the addition of distance information.  In 
either paper however, the Hubble parameter data used were not the independent 
observational measurements discussed in this review, but the ones reconstructed 
using SNIa luminosity distances.  In a similar fashion, it has been shown that 
$H(z)$ and distance measurements can further test the spatial flatness of the 
universe, or even the Copernican Principle of large-scale homogeneity and 
isotropy that is behind the mathematical form of the FRW metric 
(\ref{eq:frwmetric}) by a model-independent approach 
\citep{2008PhRvL.101a1301C,2010PhRvD..81h3537S}.  In 
\citep{2010PhRvD..81h3537S} the use of $H(z)$ in some of these tests was 
demonstrated with real-world observational data reviewed here.

Despite the wide application of the $H(z)$ datasets in the literature, we would 
like to point out some issues associated with their usage.

First, in some papers 
\citep{2010JCAP...06..002X,2010JCAP...07..018D,2010CQGra..27o5015P} that made 
use of $H(z)$ data derived from radial BAO by \citet{2009MNRAS.399.1663G} in 
$\chi^2$ analyses, the measurement at a middle redshift $z = 0.34$ was used in 
conjunction with those from the two independent redshift slices near $z = 0.24$ 
and $0.43$, under the tacit assumption of being independent from each other.  
However, this is not true, because the determination at the middle redshift was 
not made from a separate, non-overlapping redshift slice, but from the whole 
sample of galaxies, {\em including} the lower and upper redshift ranges.  If 
the data is to be used in quantitative works, this interdependency should not 
be ignored and must be explicitly analysed.  A related issue is combining the 
$H(z)$ data determined from radial BAO peaks with the $\Delta z_{\mathrm{BAO}}$ 
data derived using the same method under the assumption of their independence 
(this practice can be found, for example, in \citep{2010CQGra..27o5015P}). To 
be rigorous (or pedantic, depending on your point of view), we do not believe 
that this is the best way to use the data, and we insist on an analysis 
involving the (non-diagonal) covariance between these datasets.  On the other 
hand, the combination of $\Delta z_{\mathrm{BAO}}$ data and {\em age}-dated 
$H(z)$ is mostly free from this interdependence problem, and they actually 
complement each other well \citep[][in particular Figures 1 and 
2]{2010PhLB..689....8Z}.  We also note that in qualitative explorations one may 
choose to relax this restriction to some reasonable extent, for example in the 
discussion of accelerate expansion in \citep[Section 4.4]{2009MNRAS.399.1663G}.

\begin{figure}
    \includegraphics{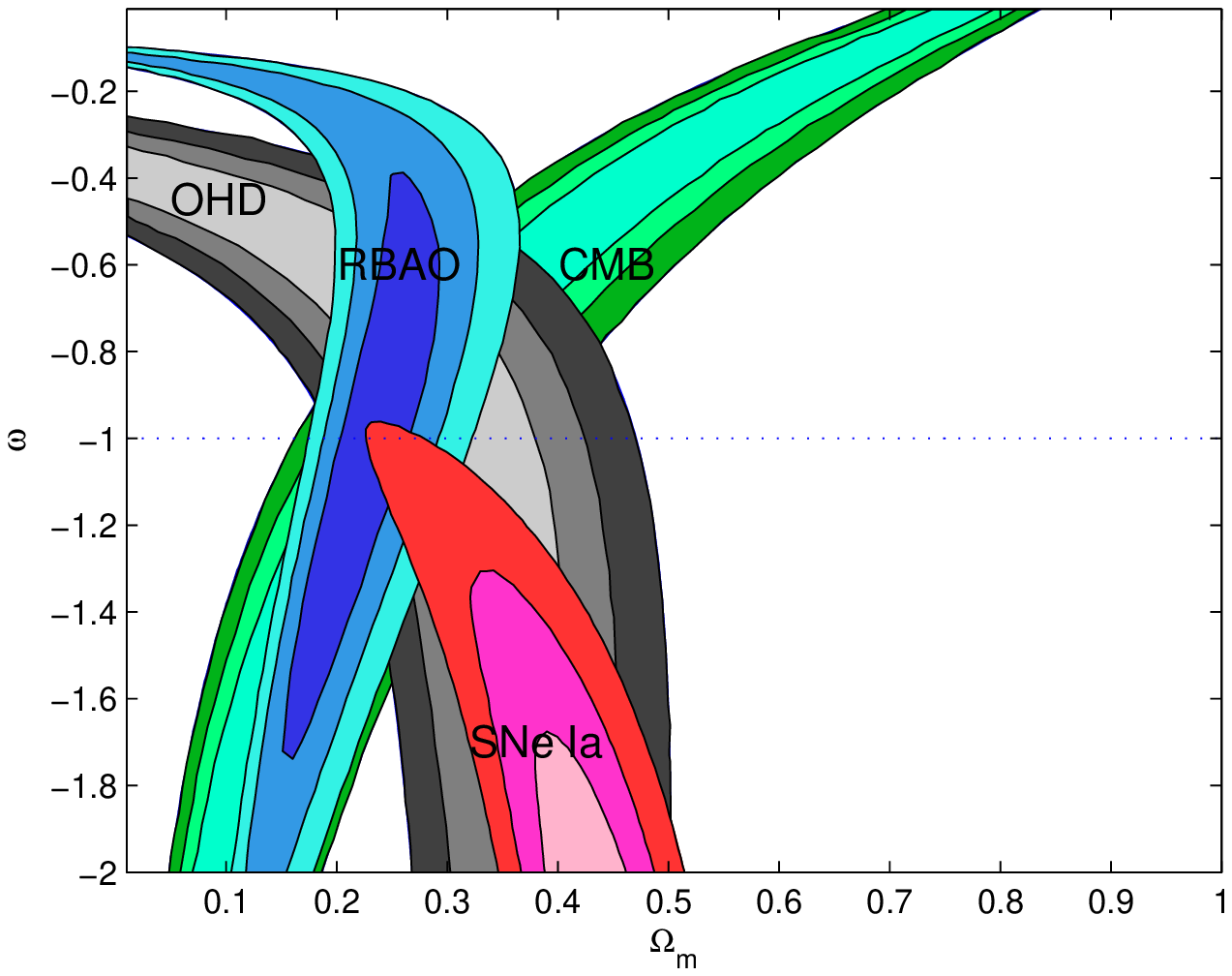}
    \caption{\label{fig:tension}Possible tension between $H(z)$ and type Ia 
    supernovae data depicted in the $\chi^2$ fitting results for the spatially 
    flat XCDM model (similar to $\lcdm$, except that the dark energy equation 
    of state parameter $\omega$ is set free instead of being fixed at 
    $\omega=-1$).  The SN data favor a phantom dark energy with $\omega < -1$ 
    while other data, including observational $H(z)$ (OHD), are consistent with 
    $\lcdm$.   The OHD used in this figure were the measurements by 
    \citep{2005PhRvD..71l3001S} using the differential age method, and the SN 
    data were from \citep{2004ApJ...607..665R}.  The RBAO contours were found 
    using the $\Delta z_{\mathrm{BAO}}$ data in \citep{2009MNRAS.399.1663G}.  
    Confidence regions are $68.3\%$, $95.4\%$, and $99.7\%$ respectively.  This 
    figure first appeared in \citep[Fig.~4]{2010PhLB..689....8Z}.}
\end{figure}

Another topic that cold be worthy of future discussions is the possible tension 
between the $H(z)$ datasets and other observational data.  As noted by 
\citet{2008JCAP...10..038F}, datasets of different physical natures and 
systematic effects can be safely combined only if they agree with each other 
well (see also \citep{2010LNP...800..147V}).  In this regard, we note that 
there is possibly some tension between $H(z)$ and type Ia supernova (SNIa) 
luminosity distances as shown in \citep{2010PhLB..689....8Z} (see Figure 
\ref{fig:tension}).  However, this apparent tension could be statistical in 
nature and may simply be a consequence of not having enough independent 
measurements of $H(z)$.  We hope that future expanded $H(z)$ datasets would 
allow us to check its consistency with other data in a quantitative manner.

\section{Future Directions}\label{sec:future}

The available $H(z)$ data have so far proven to be a useful tool in the pursuit 
of understanding the expansion history of the universe and the possible nature 
of dark energy.  However, these datasets do not have very good redshift 
coverage.  The current measurements have gone as deep as $z = 1.75$, and this 
redshift range is only sparsely covered.  There is also another issue of the 
large error bars associated with the $H(z)$ figures from the differential age 
method.  On the other hand, the collection of more and higher quality $H(z)$ 
data will not only help us constrain the parameters, but will also allow us to 
understand the possible tension between $H(z)$ and other cosmological tests.  
The latter is important, because tension is usually an indicator of systematic 
errors in the data.  By understanding the tension, we may finally conquer the 
systematic effects that have not yet been modelled well enough.

In this section, we will describe a few directions of future cosmological 
observations and their implications in the measurements of the Hubble 
parameter.

\subsection{Future Improvements in the Differential Age Method}

The relatively large uncertainties in the differential age method could be 
partially compensated if future datasets could offer better coverage in the 
redshift range accessible by this method.  Using mock data, we recently 
estimated that future $H(z)$ datasets would offer similar or even higher 
parameter-constraining power compared with current SNIa datasets if it could 
add as many as $\sim$60 independent measurements to cover the redshift range $0 
\le z \le 2$ \citep{2010arXiv1007.3787M}.  To achieve this level of data 
coverage, future surveys must be able to offer a large sample of LRGs to be 
used in age-dating.  According to \citep{2005PhRvD..71l3001S}, the Atacama 
Cosmology Telescope (ACT) 
\footnote{\url{http://www.physics.princeton.edu/act/index.html}} can be 
utilized in the future to identify passively evolving, red galaxies by their 
Sunyaev-Zel'dovich effect.  These galaxies can in turn be spectroscopically 
measured and age-dated, and it has been estimated that they could yield 
$\sim$1000 $H(z)$ measurements.  This means the quality of current differential 
age $H(z)$ data can be expected to increase significantly.

The error model used in the analysis of differential age $H(z)$ data in 
\citep{2010arXiv1007.3787M} was empirical, which may have underestimated 
possible future improvements.  In \citep{2010MNRAS.406.2569C} it has been 
estimated that $H(z)$ may be measured within $3\%$ relative error at $z \approx 
0.42$ in realistic observations if the star formation systematics could be 
properly accounted for.  This level of precision is on par with the current 
status of the radial BAO method, and we hope it could be achieved in the near 
future.

\subsection{Future Improvements in the Radial BAO Size Method}

The radial BAO size method has already been demonstrated to provide highly 
accurate $H(z)$ measurements.  However, this accuracy came at a cost, for 
spectroscopic data must be taken for the great number of galaxies under survey 
to find their redshifts, which is time-consuming.  Fortunately it turns out 
that for low redshift ranges, photometric redshift surveys can be a sufficient 
and promising approach 
\citep{2009ApJ...691..241B,2009MNRAS.394.1631A,2009JCAP...04..008R} to the 
detection and measurements of radial BAO features in the autocorrelation 
function.  Photometry has several advantages over spectroscopy --  it is 
cheaper, faster, and able to reach fainter sources.

Shortly before this review is written, the WiggleZ redshift survey 
\footnote{\url{http://wigglez.swin.edu.au/}} of emission-line galaxies produced 
its first data release \citep{2010MNRAS.401.1429D}.  As the data is being 
released, it is expected that the radial BAO signal can be put to further 
scrutiny \citep{2010ApJ...719.1032K}.

The BAO method is unique in that it allows us to reconstruct the cosmic 
expansion through a vast range of eras.  Unlike the differential age method in 
which the observable indicators of time are located within a limited redshift 
range, BAO signal detection is possible as long as the distribution of matter, 
regardless of its form, can be traced.  Even if the current implementation of 
the radial BAO method is mainly confined in the redshift range of $z \approx 
0$, future redshift surveys such as the planned SDSS III project 
\footnote{\url{http://www.sdss3.org/}} are designed to reach into deeper 
universe and measure $H(z)$ at redshifts up to $z \approx 2.5$ by observing the 
Lyman-$\alpha$ forest absorption spectra of high-redshift quasars (see 
\citep{2007PhRvD..76f3009M} for a discussion of high-$z$ measurement of radial 
BAO and $H(z)$ and its implication for dark energy, and 
\citep{2009JPhCS.180a2021N,2010ApJ...713..383W} for numerical simulation 
studies).  Recently, in the wake of the proposed Euclid satellite project 
\citep{2009ExA....23...39C,2009arXiv0912.0914L}, the enormous potential of 
space-based redshift surveys in the determination of $H(z)$ and other 
parameters has been studied in \citep{2010MNRAS.tmp.1583W}.  Finally, the 
proposed observational programs of the 21 cm background may further extend our 
knowledge of $H(z)$ into even deeper redshift ranges before or near the 
reionization era 
\citep{2005MNRAS.363L..36B,2008ApJ...673L.107M,2010ApJ...721..164S}, the ``dark 
ages'' that have not been extensively explored by current observations yet.

It is also worth noting that the previous works on the analysis and measurement 
of $H(z)$ from the clustering of LSS have mostly concentrated on the BAO 
features alone.  However, \citet{2009ApJ...693.1404S} shows that accurate 
estimates of $H(z)$ and $D_A(z)$ could be made using the full galaxy power 
spectrum in the extraction of cosmological information instead of BAO features 
alone, provided that the non-linear clustering effects are well controlled.  We 
hope that the future redshift surveys observations, as well as advances in 
better understanding of nonlinear-regime redshift-space distortions, could lead 
to successful realization of their method.

\section{Summary}\label{sec:summary}

In this paper, we reviewed the current status of observationally measured 
Hubble parameter data.  We presented the principle ideas behind the two 
important and independent methods of $H(z)$ measurement, namely the 
differential age method and the radial BAO size method.  Both methods have been 
successfully implemented over the years to yield $H(z)$ data that are of 
varying precision and redshift coverage, and the up-to-date results have been 
summarized in Table \ref{tab:obsdata}.  These data are valuable for the study 
of the expanding universe.  They have seen wide application by cosmologists to 
put various cosmological models under test, and to constrain important 
cosmological parameters either independently or in conjunction with data of 
different physical natures.  However, we also pointed out several issues in the 
usage of observational $H(z)$ data.  Finally, despite some current 
shortcomings, we find the $H(z)$ data of great potential, as future 
observational programs can be expected to improve significantly the quality of 
$H(z)$ data that may lead us into unexplored realms of the universe.

\begin{acknowledgments}

We gratefully acknowledge Chris Clarkson, Eyal A.~Kazin, and Varun Sahni for 
their helpful suggestions.  We would like to thank the anonymous referee for 
critically reviewing the manuscript and providing insightful comments that 
helped us improve this paper greatly.  CM thanks Zhongfu Yu for his help in 
preparing some of the materials in the bibliography list.  This work was 
supported by the National Science Foundation of China (Grants No.~10473002), 
the Ministry of Science and Technology National Basic Science program (project 
973) under grant No.~2009CB24901, the Fundamental Research Funds for the 
Central Universities.

\end{acknowledgments}

\bibliography{Hubcos}

\end{document}